\documentclass[twocolumn,prb,showpacs]{revtex4}
\usepackage{graphicx}
\usepackage{dcolumn}
\usepackage{bm}
\usepackage{amsmath}

\begin{document}

\preprint{}
\title{ Current induced magnetization reversal on the surface of a topological insulator }
\author{Takehito Yokoyama}
\affiliation{Department of Physics, Tokyo Institute of Technology, Tokyo 152-8551, Japan}
\date{\today}

\begin{abstract}
We study dynamics of the magnetization coupled to the surface Dirac fermions of a three dimensional topological insulator. By solving the Landau-Lifshitz-Gilbert equation in the presence of charge current, we find current induced magnetization dynamics and  discuss the possibility of magnetization reversal. The torque from the current injection depends on the transmission probability through the ferromagnet and shows nontrivial dependence on the exchange coupling. 
The magnetization dynamics is a direct manifestation of the inverse spin-galvanic effect and hence another ferromagnet is unnecessary to induce spin transfer torque in contrast to the conventional setup.

\end{abstract}

\maketitle


Topological insulator provides a new state of matter topologically
distinct from the conventional band insulator~\cite{Hasan}. In particular, edge channels or surface states are described by Dirac fermions and
protected by the band gap in  bulk states. 
Reflecting the topological nature or the surface Dirac fermion, a number of interesting phenomena have been predicted such as the quantized
magneto-electric effect~\cite{Qi,Qi2}, giant spin rotation~\cite{Yokoyama1},
magnetic properties of the surface state~\cite{Liu}, magnetization dynamics~\cite{Garate,Yokoyama3}, magneto-transport phenomena~\cite{Yokoyama2,Garate2,Mondal,Yokoyama4}, and superconducting proximity effect with Majorana fermions\cite{Fu,Fu2,Akhmerov,Tanaka,Linder}.
In particular,  from the viewpoint of spintronics, the topological insulator is an important material to pursue novel functionality because spin and momentum are tightly related on its surface.
In this paper, we utilize this  property to generate spin transfer torque. 
 
Spin transfer torque is a fundamental effect in spintronics.\cite{Tatara,Ralph} When a spin current is injected into a ferromagnetic layer with a magnetization misaligned compared to the polarization of the spin current, 
the spin angular momentum of the injected electron changes upon entering and propagating through the ferromagnetic region since it follows the direction of the magnetization via exchange coupling. This process gives a torque on the magnetization, and is called spin transfer torque. 
Consequently, the magnetization can precess \cite{Slonczewski,Tsoi} or even be switched\cite{Slonczewski2,Berger,Myers,Katine}. 
Electrical spin injection has been routinely achieved by driving a current through ferromagnet with the use of  ferromagnet/ferromagnet junction.

In this paper, 
we investigate dynamics of the magnetization coupled to the surface Dirac fermions of a three dimensional topological insulator. By solving the Landau-Lifshitz-Gilbert equation in the presence of charge current, we find current induced dynamics of the magnetization and discuss the possibility of magnetization reversal. We show that when damping is strong and the current induced torque overcomes the anisotropy field, magnetization reversal occurs. The torque from the current injection depends on the transmission probability through the ferromagnet and shows nontrivial dependence on the exchange coupling. 
The magnetization dynamics is a direct manifestation of the inverse spin-galvanic effect and hence another ferromagnet is unnecessary to induce spin transfer torque in contrast to the conventional setup.

\begin{figure}[tbp]
\begin{center}
\scalebox{0.8}{
\includegraphics[width=9.0cm,clip]{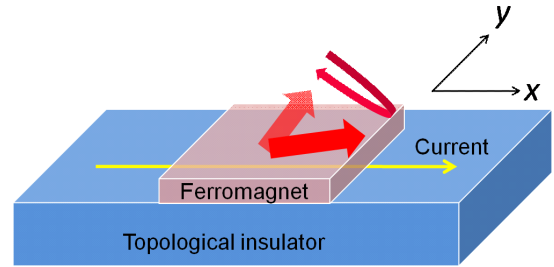}
}
\end{center}
\caption{(Color online) Schematic of the model. By injecting charge current on the surface, one can induce magnetization dynamics in the ferromagnet deposited on top of a topological insulator.}
\label{fig1}
\end{figure}


We consider magnetization dynamics of the ferromagnet deposited on the surface of a topological insulator driven by ac charge current flowing on the surface, as shown in Fig. \ref{fig1}.  
Before proceeding to the explicit calculation, let us explain the mechanism of the current induced magnetization dynamics in this setup. 
Since spin and momentum are coupled on the surface of the topological insulator,  charge current can induce the magnetization, which is the so-called the inverse spin-galvanic effect.\cite{Garate,Yokoyama3} The emergence of the inverse spin-galvanic effect is a direct consequence of the fact that,  on the surface of topological insulator, the velocity operator is given by the Pauli matrices in spin space. The current induced magnetization on the surface exerts torque on the magnetization of the ferromagnet when they are noncollinear. 
Therefore, by injecting electric current on the surface of the topological insulator, one obtains spin transfer torque between the surface Dirac fermion and the magnetization. 
 In sharp contrast, in conventional setup, another ferromagnet is necessary to inject spin polarized current into the ferromagnet to realize magnetization dynamics.

Now, let us explain the model to describe the current induced magnetization dynamics on the surface of topological insulator. 
The dynamics of the magnetization is determined by the Landau-Lifshitz-Gilbert equation:
\begin{equation}
{\dot{\bf {n}}} =  - \frac{D}{\hbar }{\bf{n}} \times \hat x + \alpha _G {\bf{n}} \times {\dot{\bf {n}}} + \frac{1}{\hbar }{\bf{T}}
\end{equation}%
where ${\bf {n}}$ is a unit vector pointing in the direction of the magnetization, $D$ represents the anisotropy energy with easy axis along $x$ direction, $\alpha _G$ is the Gilbert damping constant, and ${\bf{T}}$ is the torque acting on the magnetization. 
We consider the exchange coupling between the magnetization and the surface Dirac fermion of the form $H_{ex}={\bf{m}} \cdot {\bm{\sigma }}$ with ${\bf{m}} = h{\bf{n}}$ and the vector of the Pauli matrices in spin space ${\bm{\sigma }}$. The torque resulting from this  exchange coupling  is then given by 
\begin{eqnarray}
{\bf{T}} = h\left\langle {\bm{\sigma }} \right\rangle  \times {\bf{n}} \label{torque} .
\end{eqnarray}

Since the velocity operator is given by the Pauli matrices on the surface of a topological insulator, the expectation value of the  Pauli matrices can be represented by the charge current which in turn can be obtained through the Landauer formula: 
\begin{eqnarray}
 \left\langle {\sigma _x } \right\rangle  = \frac{{j_y }}{{ - ev_F }} = \frac{{EeV}}{{2\pi (\hbar v_F )^2 }}\int_{ - \frac{\pi }{2}}^{\frac{\pi }{2}} {d\theta \left| t \right|^2 \sin \theta }  \\ 
 \left\langle {\sigma _y } \right\rangle  = \frac{{j_x }}{{ev_F }} =  - \frac{{EeV}}{{2\pi (\hbar v_F )^2 }}\int_{ - \frac{\pi }{2}}^{\frac{\pi }{2}} {d\theta \left| t \right|^2 \cos \theta } 
\end{eqnarray}
and $ \left\langle {\sigma _z } \right\rangle =0$ where $j$, $-e$, $v_F$, $E$, and $\theta$ are the current density, the electron charge, the Fermi velocity, the Fermi energy, and the angle of incidence, respectively. Also, $t$ is transmission coefficient through the ferromagnetic region and $V$ represents applied ac voltage given by $V = V_0 \cos \Omega t$. It should be noted that transmission coefficient $t$ depends on the magnetization vector ${\bf{m}}$. 
In this way, the problem reduces to the calculation of the transmission coefficient. 

To calculate the transmission coefficient, let us explicitly write down the Hamiltonian of the system shown in Fig. \ref{fig1}: 
\begin{eqnarray}
H = \hbar v_F \left( {k_y \sigma_x  - k_x \sigma_y } \right) + {\bf{m}} \cdot {\bm{\sigma }}\Theta (x) \Theta (L - x)
\end{eqnarray}
where $\Theta (x)$ is the step function and the ferromagnet is attached to the topological insulator over the region $0<x<L$.  The magnetization of the ferromagnet is assumed to be spatially uniform. 
The wavefunctions in each region can be written as 
\begin{widetext}
\begin{eqnarray}
 \psi (x \le 0) = \frac{1}{{\sqrt 2 }}e^{ik_F x\cos \theta } \left( {\begin{array}{*{20}c}
   {ie^{ - i\theta } }  \\
   1  \\
\end{array}} \right) + \frac{r}{{\sqrt 2 }}e^{ - ik_F x\cos \theta } \left( {\begin{array}{*{20}c}
   { - ie^{i\theta } }  \\
   1  \\
\end{array}} \right) \\ 
 \psi (0 < x < L) = \frac{a}{{\sqrt {2E(E - m_z )} }}e^{i(\tilde k_x  + m_y )x} \left( {\begin{array}{*{20}c}
   {\hbar v_F (\tilde k_y  + i\tilde k_x )}  \\
   {E - m_z }  \\
\end{array}} \right) 
  + \frac{b}{{\sqrt {2E(E - m_z )} }}e^{i( - \tilde k_x  + m_y )x} \left( {\begin{array}{*{20}c}
   {\hbar v_F (\tilde k_y  - i\tilde k_x )}  \\
   {E - m_z }  \\
\end{array}} \right) \\ 
 \psi (x \ge L) = \frac{t}{{\sqrt 2 }}e^{ik_F x\cos \theta } \left( {\begin{array}{*{20}c}
   {ie^{ - i\theta } }  \\
   1  \\
\end{array}} \right)
\end{eqnarray}
\end{widetext}
where $r$ is the reflection coefficient, $E=\hbar v_F k_F$, $k_x  = k_F \cos \theta, k_y  = k_F \sin \theta$, $\hbar v_F \tilde k_x  = \sqrt {E^2  - m_z^2  - (\hbar v_F \tilde k_y^{} )^2 }$, and $\hbar v_F \tilde k_y  = \hbar v_F k_y  + m_x$.
Due to the translational invariance along the $y$-axis, the momentum $k_y$ is conserved. Hence, the common factor $e^{ik_y y}$ is omitted above.

By matching the wavefunctions at the interface $x = 0$ and $L$, we obtain the transmission coefficient $t$:
\begin{eqnarray}
t = \frac{{ - 4\cos \theta \hbar v_F \tilde k_x}}{{\alpha (A + ie^{i\theta } B)}}
\end{eqnarray}
where
\begin{widetext}
\begin{eqnarray}
 A = \left[ {\alpha _2 \left\{ {ie^{ - i\theta } \hbar v_F (\tilde k_y  + i\tilde k_x ) - E - m_z } \right\} - \alpha _1 \left\{ {ie^{ - i\theta } \hbar v_F (\tilde k_y  - i\tilde k_x ) - E - m_z } \right\}} \right], \\ 
 B = \left[ {\alpha _2 \left\{ {ie^{ - i\theta } (E - m_z ) - \hbar v_F (\tilde k_y  - i\tilde k_x )} \right\} - \alpha _1 \left\{ {ie^{ - i\theta } (E - m_z ) - \hbar v_F (\tilde k_y  + i\tilde k_x )} \right\}} \right],
\end{eqnarray}
\end{widetext}
$\alpha  = e^{ik_F L\cos \theta } ,\alpha _1  = e^{i(\tilde k_x  + m_y )L}$, and $\alpha _2  = e^{i( - \tilde k_x  + m_y )L}$.

By using the expression of the transmission coefficient, we obtain the torque steming from the current injection  Eq.(\ref{torque}), and finally we solve the Landau-Lifshitz-Gilbert equation numerically. 
In the following, the initial condition is set to be $n_x(t=0)=1, n_y(t=0)=0$ and  $n_z(t=0)=0$.  We set $\omega _F  = D/\hbar$ and also introduce a parameter $\bar V = \frac{{E^2 eV_0 }}{{2\pi \hbar ^3 \omega _F v_F^2 }}$ which measures the  applied voltage relative to the anisotropy energy since $\bar V \propto \frac{{eV_0 }}{D}$.

\begin{figure}[tbp]
\begin{center}
\scalebox{0.8}{
\includegraphics[width=7.50cm,clip]{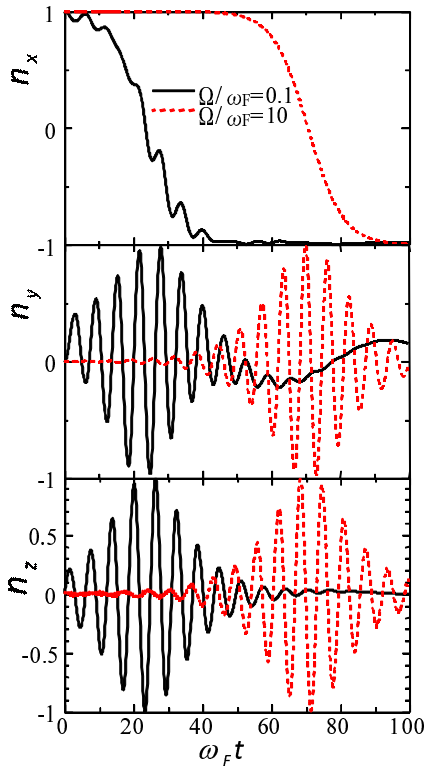}
}
\end{center}
\caption{(Color online) Time evolution of the magnetization vector with $h/E=0.1$,  $\bar V = 1$, $k_F L=100$, and $\alpha _G =0.1$. Solid line $\Omega/\omega_F=0.1$. Dotted line $\Omega/\omega_F=10$.}
\label{fig2}
\end{figure}

\begin{figure}[tbp]
\begin{center}
\scalebox{0.8}{
\includegraphics[width=7.50cm,clip]{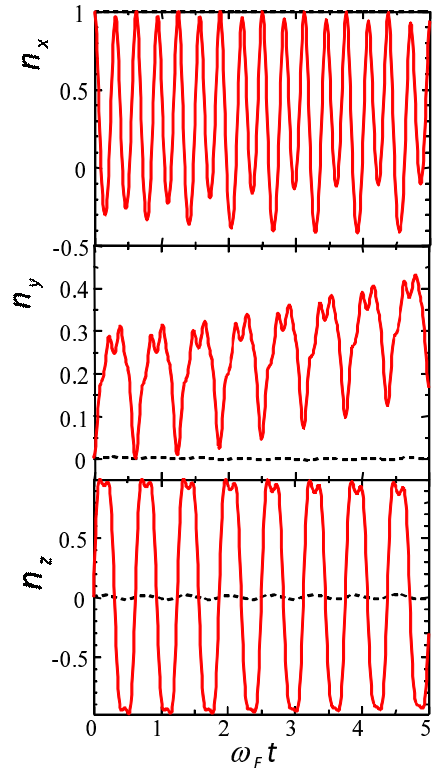}
}
\end{center}
\caption{(Color online) Time evolution of the magnetization vector for weaker damping $\alpha _G =0.01$, $h/E=0.1$, $k_F L=100$, and $\Omega/\omega_F=10$. Solid line $\bar V = 100$.  Dotted line  $\bar V = 1$.}
\label{fig3}
\end{figure}

\begin{figure}[tbp]
\begin{center}
\scalebox{0.8}{
\includegraphics[width=8.0cm,clip]{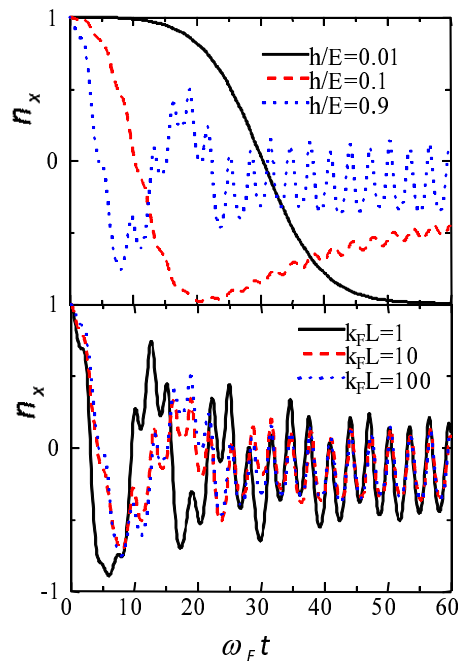}
}
\end{center}
\caption{(Color online) Time evolution of $n_x$ for $\alpha _G =0.1$, $\bar V = 1$ and $\Omega/\omega_F=1$ with $k_F L=100$ and  various $h/E$ (upper), and with $h/E=0.9$ and various $k_F L$ (lower).}
\label{fig4}
\end{figure}

In Fig. \ref{fig2}, we show time evolution of the magnetization vector for $\Omega/\omega_F=0.1$ and $10$ with $h/E=0.1$, $\bar V = 1$, $k_F L=100$, and  $\alpha _G =0.1$. We find the current induced magnetization reversal from $n_x=1$ to $n_x=-1$. When $n_x$ is reversed, the other components $n_y$ and $n_z$ strongly oscillate. As $\Omega$ increases, the magnetization reversal occurs at later time. This is because, for large $\Omega$, the torque from the charge current oscillates very fast compared to the characteristic motion of the magnetization. Hence, the magnetization would experience an averaged torque over many oscillations, which results in small effect due to partial cancellation of the net torque.

Figure \ref{fig3} shows the time evolution of the magnetization vector for smaller damping constant $\alpha _G =0.01$ with $h/E=0.1$, $k_F L=100$ and $\Omega/\omega_F=10$. At  $\bar V = 1$,  the magnetization vector is trapped near the initial condition and the dynamics is quite weak (see dotted line). To induce stronger dynamics, one should increase the torque arising from the current injection.  By increasing the bias voltage ($\bar V = 100$), the dynamics becomes stronger and we find strong oscillation of the magnetization, including  reversal of the magnetization but the reversal is not permanent due to a small damping constant as shown by solid line in Fig. \ref{fig3}. 

One may think that for larger exchange coupling, the oscillation of the magnetization becomes stronger since the torque (Eq.(\ref{torque})) seems to become larger. However,  when the value of $m_x$ becomes large, the Fermi surface moves in the $k_y$-direction in the ferromagnetic region\cite{Yokoyama2}. Then, the transmission probability becomes very small because the number of the evanescent modes increases. Therefore, in effect, for large $h$, the torque does not contribute when the magnetization vector points to $x$-direction. 
Since $\left\langle {\bm{\sigma }} \right\rangle$ lies in the $x-y$ plane, the torque (Eq.(\ref{torque})) represents an easy axis anisotropy along the axis parallel to $\left\langle {\bm{\sigma }} \right\rangle$ in $x-y$ plane \cite{Garate}, but when the magnetization vector points to $x$-direction, the torque becomes very small. Thus, for large $h$, the torque
virtually represents an easy axis anistropy along $y$-axis. 
This indicates the deviation of the magnetization from the $x$-axis for large $h$. 
We show time evolution of  $n_x$  for $\alpha _G =0.1$, $\bar V = 1$ and $\Omega/\omega_F=1$  with $k_F L=100$ and  various $h/E$ in  the upper panel of Fig. \ref{fig4}, and with $h/E=0.9$ and various $k_F L$ in the lower panel of Fig. \ref{fig4}. 
Consistent with the above discussion, the magnitude of $n_x$ becomes small for large $h$, while other components, $n_y$ and $n_z$, become large. Thus, we find that optimal magnitude of the exchange coupling for the magnetization reversal is in the intermediate regime of the exchange coupling (note that when $h \to 0$, no current induced dynamics occurs since ${\bf{T}} \to {\bf{0}}$). 
As $L$ increases, the transmission probability decreases and hence the torque from the current injection is suppressed. Thus, the magnitude of the oscillation decreases with increasing $L$ as shown in the lower panel of Fig. \ref{fig4}.
However, for smaller $h$, since the number of the evanescent modes decreases, the effect of the length of the ferromagnet $L$ becomes less prominent.

As for experimental realizability, 
if we set $E=$100 meV and $k_F=1$ nm$^{-1}$, then the parameter range used in this paper is $L=1 \sim 100$ nm and $h = 1 \sim 10$ meV. Typically, $\hbar \omega_F=D \simeq $ 0.1 meV, then $\bar V=1$ corresponds to $eV_0 \simeq  0.1$ meV and the value of $\Omega$ used in this paper is in  THz regime. These values can be achieved by the present experimental techniques. 

In summary,
we have studied  dynamics of the magnetization  coupled to the surface Dirac fermions of a three dimensional topological insulator. By solving the Landau-Lifshitz-Gilbert equation in the presence of charge current, we have found current induced dynamics of the magnetization and discussed the possibility of magnetization reversal. The torque from the current injection depends on the transmission probability through the ferromagnet and shows nontrivial dependence on the exchange coupling. 
The magnetization dynamics is a direct manifestation of the inverse spin-galvanic effect and hence another ferromagnet is unnecessary to induce spin transfer torque in contrast to the conventional setup.

This work was supported by Grant-in-Aid for Young Scientists (B) (No. 23740236) and the "Topological Quantum Phenomena" (No. 23103505) Grant-in Aid for Scientific Research on Innovative Areas from the Ministry of Education, Culture, Sports, Science and Technology (MEXT) of Japan.


\begin{thebibliography}{99}

\bibitem{Hasan} X. L. Qi and S. C. Zhang, Physics Today, \textbf{63}, 33 (2010); M. Z. Hasan and C. L. Kane, Rev. Mod. Phys. \textbf{82}, 3045, (2010); X. L. Qi and S. C. Zhang, arXiv:1008.2026v1.

\bibitem{Qi} X.-L. Qi, T. Hughes, and S.-C. Zhang, Nature Phys. \textbf{4}, 273 (2008).

\bibitem{Qi2} X.-L. Qi, T. L. Hughes, and S.-C. Zhang, Phys. Rev. B \textbf{78},
195424 (2008).

\bibitem{Yokoyama1} T. Yokoyama, Y. Tanaka, and N. Nagaosa, Phys. Rev. Lett.
\textbf{102}, 166801 (2009).

\bibitem{Liu} Q. Liu, Chao-Xing Liu, C. Xu, Xiao-Liang Qi, and Shou-Cheng Zhang, Phys. Rev. Lett. \textbf{102}, 156603 (2009). 

\bibitem{Garate} I. Garate and M. Franz, Phys. Rev. Lett. \textbf{104}, 146802 (2010).

\bibitem{Yokoyama3} T. Yokoyama, J. Zang, and N. Nagaosa, Phys. Rev. B \textbf{81}, 241410(R) (2010). 

\bibitem{Yokoyama2} T. Yokoyama, Y. Tanaka, and N. Nagaosa, Phys. Rev. B \textbf{81}, 121401(R) (2010).

\bibitem{Garate2} I. Garate and M. Franz, Phys. Rev. B \textbf{81}, 172408 (2010).

\bibitem{Mondal} S. Mondal, D. Sen, K. Sengupta, and R. Shankar, Phys. Rev. Lett. \textbf{104}, 046403 (2010).

\bibitem{Yokoyama4} T. Yokoyama and S. Murakami, Phys. Rev. B \textbf{83}, 161407(R) (2011). 


\bibitem{Fu} L. Fu and C. L. Kane, Phys. Rev. Lett. \textbf{100}, 096407
(2008).

\bibitem{Fu2} L. Fu and C. L. Kane, Phys. Rev. Lett. \textbf{102}, 216403 (2009). 

\bibitem{Akhmerov} A. R. Akhmerov, J. Nilsson, and C. W. J. Beenakker, Phys. Rev. Lett. \textbf{102}, 216404 (2009). 

\bibitem{Tanaka} Y. Tanaka, T. Yokoyama, and N. Nagaosa, Phys. Rev. Lett.
\textbf{103}, 107002 (2009).

\bibitem{Linder} J. Linder, Y. Tanaka, T. Yokoyama, A. Sudb{\o}, and N. Nagaosa, Phys. Rev. Lett. \textbf{104}, 067001 (2010);Phys. Rev. B \textbf{81}, 184525 (2010). 


\bibitem{Tatara} G. Tatara, H. Kohno, and J. Shibata, Phys. Rep. \textbf{468}, 213 (2008). 

\bibitem{Ralph} D. C. Ralph and M. D. Stiles, J. Magn. Magn. Mater. \textbf{320}, 1190 (2008).

\bibitem{Slonczewski} J. C. Slonczewski, J. Magn. Magn. Mater. \textbf{195}, L261 (1999). 

\bibitem{Tsoi} M. Tsoi, A. G. M. Jansen, J. Bass, W.-C. Chiang, M. Seck, V.
Tsoi, and P. Wyder, Phys. Rev. Lett. \textbf{80}, 4281 (1998); M. Tsoi,
A. G. M. Jansen, J. Bass, W.-C. Chiang, V. Tsoi, and P. Wyder,
Nature \textbf{406}, 46 (2000).

\bibitem{Slonczewski2} J. Slonczewski, J. Magn. Magn. Mater. \textbf{159}, L1 (1996).

\bibitem{Berger} L. Berger, Phys. Rev. B \textbf{54}, 9353 (1996).

\bibitem{Myers} E. B. Myers, D. C. Ralph, J. A. Katine, R. N. Louie, and R. A.
Buhrman, Science \textbf{285}, 867 (1999).

\bibitem{Katine} J. A. Katine, F. J. Albert, R. A. Buhrman, E. B. Myers, and D. C. Ralph, Phys. Rev. Lett. \textbf{84}, 3149 (2000).



\end{thebibliography}
\end{document}